\documentclass[twocolumn,showpacs,preprintnumbers,amsmath,amssymb,prl]{revtex4}
\usepackage{dcolumn}
\usepackage{bm}
\usepackage[dvips]{graphicx}
\usepackage{wrapfig}
\usepackage{psfrag}
\begin{document}

\def\sh{\mathop{\rm sh}\nolimits}
\def\ch{\mathop{\rm ch}\nolimits}
\def\var{\mathop{\rm var}}\def\exp{\mathop{\rm exp}\nolimits}
\def\Re{\mathop{\rm Re}\nolimits}
\def\Sp{\mathop{\rm Sp}\nolimits}
\def\kp{\mathop{\text{\ae}}\nolimits}
\def\bk{{\bf {k}}}
\def\bp{{\bf {p}}}
\def\bq{{\bf {q}}}
\def\lra{\mathop{\longrightarrow}}
\def\Const{\mathop{\rm Const}\nolimits}
\def\sh{\mathop{\rm sh}\nolimits}
\def\ch{\mathop{\rm ch}\nolimits}
\def\var{\mathop{\rm var}}
\def\mK{\mathop{{\mathfrak {K}}}\nolimits}
\def\mR{\mathop{{\mathfrak {R}}}\nolimits}
\def\mv{\mathop{{\mathfrak {v}}}\nolimits}
\def\mV{\mathop{{\mathfrak {V}}}\nolimits}
\def\mD{\mathop{{\mathfrak {D}}}\nolimits}
\def\mN{\mathop{{\mathfrak {N}}}\nolimits}
\def\mS{\mathop{{\mathfrak {S}}}\nolimits}

\def\Re{\mbox {Re}}
\newcommand{\Z}{\mathbb{Z}}
\newcommand{\R}{\mathbb{R}}
\def\mK{\mathop{{\mathfrak {K}}}\nolimits}
\def\mR{\mathop{{\mathfrak {R}}}\nolimits}
\def\mv{\mathop{{\mathfrak {v}}}\nolimits}
\def\mV{\mathop{{\mathfrak {V}}}\nolimits}
\def\mD{\mathop{{\mathfrak {D}}}\nolimits}
\def\mN{\mathop{{\mathfrak {N}}}\nolimits}
\newcommand{\ccm}{{\cal M}}
\newcommand{\cE}{{\cal E}}
\newcommand{\cV}{{\cal V}}
\newcommand{\cI}{{\cal I}}
\newcommand{\cR}{{\cal R}}
\newcommand{\cK}{{\cal K}}
\newcommand{\cH}{{\cal H}}

\def\br{\mathop{{\bf {r}}}\nolimits}
\def\bS{\mathop{{\bf {S}}}\nolimits}
\def\bA{\mathop{{\bf {A}}}\nolimits}
\def\bJ{\mathop{{\bf {J}}}\nolimits}
\def\bn{\mathop{{\bf {n}}}\nolimits}
\def\bg{\mathop{{\bf {g}}}\nolimits}
\def\bv{\mathop{{\bf {v}}}\nolimits}
\def\be{\mathop{{\bf {e}}}\nolimits}
\def\bp{\mathop{{\bf {p}}}\nolimits}
\def\bz{\mathop{{\bf {z}}}\nolimits}
\def\bbf{\mathop{{\bf {f}}}\nolimits}
\def\bb{\mathop{{\bf {b}}}\nolimits}
\def\ba{\mathop{{\bf {a}}}\nolimits}
\def\bx{\mathop{{\bf {x}}}\nolimits}
\def\by{\mathop{{\bf {y}}}\nolimits}
\def\br{\mathop{{\bf {r}}}\nolimits}
\def\bs{\mathop{{\bf {s}}}\nolimits}
\def\bH{\mathop{{\bf {H}}}\nolimits}
\def\bk{\mathop{{\bf {k}}}\nolimits}
\def\be{\mathop{{\bf {e}}}\nolimits}
\def\bnul{\mathop{{\bf {0}}}\nolimits}
\def\bq{{\bf {q}}}

\newcommand{\oV}{\overline{V}}
\newcommand{\vkp}{\varkappa}
\newcommand{\os}{\overline{s}}
\newcommand{\opsi}{\overline{\psi}}
\newcommand{\ov}{\overline{v}}
\newcommand{\oW}{\overline{W}}
\newcommand{\oPhi}{\overline{\Phi}}

\def\mI{\mathop{{\mathfrak {I}}}\nolimits}
\def\mA{\mathop{{\mathfrak {A}}}\nolimits}

\def\st{\mathop{\rm st}\nolimits}
\def\tr{\mathop{\rm tr}\nolimits}
\def\sign{\mathop{\rm sign}\nolimits}
\def\d{\mathop{\rm d}\nolimits}
\def\const{\mathop{\rm const}\nolimits}
\def\O{\mathop{\rm O}\nolimits}
\def\Spin{\mathop{\rm Spin}\nolimits}
\def\exp{\mathop{\rm exp}\nolimits}

\def\mI{\mathop{{\mathfrak {I}}}\nolimits}
\def\mA{\mathop{{\mathfrak {A}}}\nolimits}

\def\st{\mathop{\rm st}\nolimits}
\def\tr{\mathop{\rm tr}\nolimits}
\def\sign{\mathop{\rm sign}\nolimits}
\def\d{\mathop{\rm d}\nolimits}
\def\const{\mathop{\rm const}\nolimits}
\def\O{\mathop{\rm O}\nolimits}
\def\Spin{\mathop{\rm Spin}\nolimits}
\def\exp{\mathop{\rm exp}\nolimits}

\title{Discrete quantum gravity: continuum limit
and the problem of state doubling}

\author {S.N. Vergeles\vspace*{4mm}\footnote{{e-mail:vergeles@itp.ac.ru}}}

\affiliation{Landau Institute for Theoretical Physics,
Russian Academy of Sciences,
Chernogolovka, Moskow region, 142432 Russia }

\begin{abstract} It is shown that in the theory of discrete quantum gravity
defined on the irregular "breathing" lattice,
if the macroscopic continuum phase is realized,
the phenomenon of state doubling
(even if it exists formally at kinematic level) actually is absent at
experimentally accessible energies.
\end{abstract}


\maketitle

{\bf 1.} In the recent my works one more version of the
discrete quantum gravity has been introduced into consideration
\cite{1}-\cite{5}. This version of discrete quantum gravity manifests
"naive" macroscopic continuum limit. But it does not mean that
in the theory the macroscopic continuum phase indeed is
realized dynamically. Moreover, some phenomena and problems are well known
which prevent microscopic discrete quantum gravity from developing
into macroscopic continuum phase with the properties close to the real
one. For example, the phenomenon of state doubling (Wilson doubling)
and the cosmological constant problem.

In the present and the subsequent letters I study the
state doubling phenomenon and the cosmological constant problem
under the assumption that the macroscopic continuum phase
is realized dynamically in the theory of discrete gravity of the type \cite{5}.
I think that this is a very strong assumption: for realization of
this assumption the microscopic theory must be determined
very specially and, perhaps, uniquely (the correct set of fundamental fields
and their interaction, and so on). Though all that is unknown,
I apply a sort of phenomenological approach and
study the mentioned problems in the macroscopic phase.
It seems to me that the knowledge of some qualitative properties
of the considered theories in the macroscopic phase can promote general
progress of this branch of quantum field theory.

In the present letter I study the possible consequences due to
the phenomenon of state doubling. The problem of graviton
states doubling in discrete gravity was considered earlier
in \cite{6}. The fermion state doubling (proper Wilson doubling, see,
for example \cite{7})
was investigated in \cite{1}-\cite{4}. For simplicity I
study here the Dirac field on the lattice. But it is clear
from the consideration given below that the conclusion
remains valid with respect to all other fields.

At first sight it seems that the theories in which the
state doubling phenomenon is present
(I call them as anomalous theories) must be very different from
the usual theories (without state doubling) in the continuum phase.
Indeed, in anomalous theories the transitions from the
states with normal modes into the states with anomalous modes
(and vice versa) exist. But the such transitions are not
registered at experimentally accessible energies.

The main result of the paper is the behaviour in the $x$-space
of the propagator describing the propagation of anomalous modes
\begin{gather}
\left\langle\langle\psi^{{\cal A}}(x)\opsi^{{\cal
A}}(y)\rangle\right\rangle_{e_{\cE}}\sim\exp\left[-\frac{|x-y|}{a\lambda}\right].
\label{introduction10}
\end{gather}
Here and below $a$ is a certain constant that has the length
dimension and is of the order of the effective lattice step
and $\lambda$ is the number of the order of unit.
But the propagators describing the propagation of normal modes
have the conformal behaviour at the enough large energies.
Thus, the propagators of anomalous modes decrease too fast
as compared with the propagators of normal modes. Therefore
the transitions between the states of normal and anomalous
modes are strongly depressed at the scales much greater than
$a$. The situation is similar to the following one: the
electron loop corrections to the external smooth electromagnetic
field die out at the scales much greater than $1/m$ ($m$ is the
electron mass).

{\bf 2.} It is necessary to write out some formulae from the work \cite{5}.
Here the notations are completely identical to that in \cite{5}.

Let $\mK$ be a 4-dimensional simplicial complex such that the
3-dimensional complex $\mS=\partial\mK$ has the topology of 3-sphere $S^3$.
To each vertex $a_i\in\mK$, the Dirac spinors
$\psi_i$ and $\overline{\psi}_i$ belonging to the
complex Grassman algebra are assigned.
To each oriented edge $a_ia_j\in\mK$, an element of the
group  $Spin(4)$
\begin{eqnarray}
\Omega_{ij}=\Omega^{-1}_{ji}=\exp\left(\frac{1}{2}\omega^{ab}_{ij}
\sigma^{ab}\right)\,, \ \ \ \sigma^{ab}=\frac{1}{4}[\gamma^a,\,\gamma^b]\,,
\label{discr20}
\end{eqnarray}
and also an element $\hat{e}_{ij}\equiv
e^a_{ij}\gamma^a$, such that
\begin{eqnarray}
\hat{e}_{ij}=-\Omega_{ij}\hat{e}_{ji}\Omega_{ij}^{-1}, \quad
-\infty<e^a_{ij}<\infty,
\label{discr30}
\end{eqnarray}
are assigned. The notations
$\overline{\psi}_{Ai}, \ \psi_{Ai}, \ \hat{e}_{Aij}, \
\Omega_{Aij}$ and so on indicate that the edge $X^A_{ij}=a_ia_j$ belongs to 4-simplex
with index $A$. Let $a_{Ai}, \ a_{Aj}, \ a_{Ak}, \ a_{Al}$, and $a_{Am}$ be all
five vertices of a 4-simplex with index $A$ and
$\varepsilon_{Aijklm}=\pm 1$ depending on whether the order of
vertices $a_{Ai}\,a_{Aj}\,a_{Ak}\,a_{Al}\,a_{Am}$ defines the
positive or negative orientation of this 4-simplex. The Euclidean action
of the theory has the form
\begin{eqnarray}&
I=\frac{1}{5\times\label{discr40}
24}\sum_A\sum_{i,j,k,l,m}\varepsilon_{Aijklm}\tr\,\gamma^5 \times
\\& \nonumber
\times\left\{-\frac{1}{2\,l^2_P}\Omega_{Ami}\Omega_{Aij}\Omega_{Ajm}
\hat{e}_{Amk}\hat{e}_{Aml}+\right.
\\&\!\!\!\!\!
\left.+\frac{i}{48}\gamma^a\!\!\left(\overline{\psi}_{Ai}\gamma^a
\Omega_{Aij}\psi_{Aj}-\overline{\psi}_{Aj}\Omega_{Aji}\gamma^a\psi_{Ai}\right)
\hat{e}_{Amj}\hat{e}_{Amk}\hat{e}_{Aml}\right\}\!\!\!\!\!\!
\nonumber
\end{eqnarray}

The oriented volume of a 4-simplex with vertexes
$a_{Ai}, \ a_{Aj}, \ a_{Ak}, \ a_{Al}$, and $a_{Am}$ is equal to
\begin{eqnarray}
V_A=\frac{1}{4!}\frac{1}{5!}\sum_{i,j,k,l,m}\varepsilon_{A\,ijklm}\,
\varepsilon^{abcd}\,e^a_{A\,mi}e^b_{A\,mj}e^c_{A\,mk}e^d_{A\,ml}\,.
\label{discr80}
\end{eqnarray}
The quantity
$l^2_{ij}=\sum_{a=1}^4(e^a_{ij})^2$
is interpreted as the square of the length of the edge $a_ia_j$.

The partition function $Z$ for a discrete
Euclidean gravity \footnote{I mean here that the partition
function is a functional of the values of the dynamic
variables at the boundary $\partial\mK$.} is defined as follows:
\begin{gather}
Z=\const\cdot\bigg (\prod_{\cE}\,\int\, \d\Omega_{\cE}\,\int\,\d
e_{\cE}\,\bigg)\times
\nonumber \\
\times\big(\prod_{\cV}\,\d\overline{\psi}_{\cV}\,
\d\psi_{\cV}\,\big)\,\exp\big(-I\,\big)\,.
\label{discr90}
\end{gather}
Here, the vertices and edges are enumerated
by indices $\cV$ and $\cE$, and the
corresponding variables are denoted by $\psi_{\cV}$, \ $\Omega_{\cE}$,
respectively, $\d\Omega_{\cE}$
is the Haar measure on the group $\Spin(4)$, and
\begin{gather}
\d e_{\cE}\equiv\prod_a\,\d e^a_{\cE}\,, \quad \
\d\overline{\psi}_{\cV}\,\d\psi_{\cV}\equiv\prod_{\nu}\,
\d\overline{\psi}_{\cV\nu}\,\d\psi_{\cV\nu}\,.
\label{discr100}
\end{gather}
The index $\nu$ enumerates individual components of the
spinors $\psi_{\cV}$ and $\overline{\psi}_{\cV}$.

Let's denote by ${\cal X}$ a 4-dimensional smooth manifold with topology of
the complex $\mK$. Consider a set of maps $\{g\}$ from geometrical realization
of the complex $\mK$ onto manifold ${\cal X}$ which
are not necessarily one-one maps. For a given local coordinates
$x^{\mu},\,\mu=1,\,2,\,3,\,4$ a map $g$ defines the coordinates
of images of vertexes $a_{A\,i}$: $x^{\mu}_{g(A\,i)}\equiv g^{\mu}(a_{A\,i})$.
Define the four differentials
\begin{gather}
\d x^{\mu}_{g(A\,ji)}\equiv x^{\mu}_{g(A\,i)}-x^{\mu}_{g(A\,j)}\,,
\quad i\neq j\,, \ \ \ i=1,\,\dots,4.
\label{discr140}
\end{gather}
Suppose the smooth fields $\omega_{\mu}^{ab}(x), e^a_{\mu}(x), \opsi(x),
\psi(x)$ are defined on the manifold ${\cal X}$. Then we can define
the discrete lattice variables according to the relations
\begin{gather}
\omega_{\mu}^{ab}(x_{g(A\,m)})\d x^{\mu}_{g(A\,mi)}=\omega_{A\,mi}^{ab},
\nonumber \\
e_{\mu}^a(x_{g(A\,m)})\d x^{\mu}_{g(A\,mi)}=e_{A\,mi}^a,
\quad \psi(x_{g(A\,i)})=\psi_{A\,i}.
\label{discr150}
\end{gather}
On the contrary, the discrete lattice variables in the right hand sides
of Eqs. (\ref{discr150}) define the values of the fields on the images
of vertexes of the complex. It is clear \cite{5} that for
the discrete lattice variables which change enough smoothly along the
complex we obtain the enough smooth fields. Moreover, in this case
the discrete action (\ref{discr40}) transforms to the well known
continuum action
\begin{eqnarray}&
I=\int\,\varepsilon_{abcd}\,\left\{-\frac{1}{l^2_P}R^{ab}\wedge e^c\wedge e^d+\right.
\nonumber \\&
\left.+\frac{i}{12}\,\big(\,\opsi\,\gamma^a\, {\cal
D}_{\mu}\psi-{\cal D}_{\mu}\opsi\,\gamma^a\,\psi\,\big)\,\d
x^{\mu}\wedge e^b\wedge e^c\wedge e^d\right\},
\nonumber \\&
e^a=e^a_{\mu}\d x^{\mu}, \qquad \omega^{ab}=\omega^{ab}_{\mu}\d x^{\mu},
\label{discr160}
\\&
\d\omega^{ab}+\omega^a_c\wedge\omega^{cb}=\frac{1}{2}\,R^{ab}.
\nonumber
\end{eqnarray}
I emphasize that we obtain the action (\ref{discr160}) only if the
lowest derivatives of the fields are taken into account. It is important
that in this case the information on the structure of the complex is lost.
This is incorrectly if the highest derivatives of the fields are
also taken into account. In the work \cite{5} the arguments have been given
that the quasi-classical phase at the same time is the macroscopic continuum phase with
long correlations and hence, also, the phase in which
the highest derivatives of the fields can be ignored. In this phase the partition
function is saturated by normal (smooth) modes but not by anomal modes responsible for
Wilson state doubling.

It is important that in the quasi-classical phase the universe wave function
does not depend on the discrete variables $e^a_{A\,ij}$ in a wide
diapason. This statement is true in the same context as the
highest derivatives of the fields can be ignored. Indeed, in the
quasi-classical phase the action (\ref{discr160}) as well as
the universe wave function depend on the fields
$\omega_{\mu}^{ab}(x), e^a_{\mu}(x), \opsi(x), \psi(x)$
which are present in the left hand side of Eqs. (\ref{discr150}).
Thus, fixing these fields and varying the maps $g$ or, equivalently,
the images of vertexes $x^{\mu}_{g(A\,i)}$
and, hence, the differentials $\d x^{\mu}_{g(A\,ji)}$, one can
vary the discrete variables $e_{A\,mi}^a$
in the right hand side of Eqs. (\ref{discr150}) in a wide region.

To clarify the situation, let's consider the case when
geometry of the space-time is flat or almost flat.
In the flat case one can take
\begin{gather}
\omega_{ij}^{ab}=0\,\,, \ \ \
\bigl(e_{ij}^a+e_{jk}^a+\ldots+e_{li}^a\bigr)=0\,\,.
\label{discr170}
\end{gather}
Here, the sum in the parentheses is taken over any closed path
formed by 1-simplices. Equations (\ref{discr170}) indicates that the
curvature and torsion are equal to zero. Thus, geometrical
realization of the complex $\mK$
is in the four-dimensional Euclidean space, $e^a_{ij}$ being the
components of the vector in a certain orthogonal basis in this
space, and if $R^a_i$ is the radius-vector of vertex $a_i$, then
$e_{ij}^a=R^a_j-R^a_i$. In this case one can take $e^a_{\mu}(x)=\delta^a_{\mu}$.
It is evident that Eqs. (\ref{discr170}) are the only restrictions for
the variables $e^a_{ij}$.

{\bf 3.} Let us write out the equation for the eigenmodes of the discrete
Dirac operator in partial case (\ref{discr170}). This operator
is obtained by varying action (\ref{discr40}) with respect
to the variable $\opsi_i$. In the four-dimensional case,
we have (see \cite{3})
\begin{gather}
\frac{i}{4}\sum_{j(i)}S^a_{ij}\gamma^a\psi_j=\epsilon\,v_i\psi_i,
\label{ancor310}
\end{gather}
where
\begin{gather}
S^a_{ij}=(3!)^{-2}\sum_{A(i,j)}
\nonumber \\
\times\sum_{k,l,m}\varepsilon^{acdf}
\varepsilon_{A(i,j)\,ijklm}e^c_{A(i,j)ik}e^d_{A(i,j)il}e^f_{A(i,j)im}.
\label{ancor320}
\end{gather}
Here, the index $A(i,j)$ enumerates all the 4-simplexes containing the edge
$a_ia_j$, the index $j(i)$ enumerates all the vertices $a_{j(i)}$
neighboring the vertex $a_i$ \footnote{The
vertices $a_i$ and $a_j$ are called neighboring if
the 1-simplex $a_ia_j$ exists.}, and $v_i$ is the sum of oriented 4-volumes of
all 4-simplexes containing the vertex $a_i$.
There exists the relations
\begin{gather}
\sum_{j(i)}S^a_{ij}\,e^b_{ij}=4\,v_i\,\delta^{ab}, \quad
 \sum_{j(i)}S^a_{i,j}\equiv 0,
 \label{ancor330}
\end{gather}
that provide the transformation of discrete equation (\ref{ancor310})
in the long-wavelength limit to the continuum equation
\begin{gather}
i\gamma^a\partial_a\psi=\epsilon\,\psi.
\label{ancor340}
\end{gather}
Indeed, this is easily verified by taking into account
the relation
\begin{gather}
\psi_j=\psi_i+e^a_{ij}\partial_a\psi_i,
\label{ancor350}
\end{gather}
which is almost true in the long-wavelength limit, i.e., for conventional modes.
Here, $x^a$ are the Cartesian coordinates in the same orthonormal
basis in which the components of the vectors $e^a_{ij}$ are specified.
It is again seen that information on the positions of vertices of the
complex is completely lost in the long-wavelength limit; i.e.,
conventional long-wavelength modes lose information on the
lattice structure. The eigenvalues of the modes of Dirac
operator (\ref{ancor340}) are $\epsilon=\pm |k|$, where $k^a$
is the wavevector of a mode. Thus, the eigenvalues of the usual modes
can be arbitrarily small.

The main result of works \cite{2}, \cite{3} is that the theory
under consideration involves two types of lattices. On the
lattices of one type, fermion doubling occurs, whereas it is
absent on the lattices of the other type.

According to the definition, the lattice allows fermion state
doubling if the discrete Dirac operator given by Eq. (\ref{ancor310})
allows two types of modes with eigenvalues approaching
zero\footnote{It is certainly assumed that the sizes of
the lattice (i.e. the number of its simplexes) tend to infinity}.
The qualitative difference between them is as follows.
The normalized modes of the first type, or normal modes,
satisfy the conditions
\begin{gather}
|\psi_i-\psi_j|\sim|\epsilon|\,|\psi_j|\sim|\epsilon|\,N^{-1/2},
\nonumber
\end{gather}
where $a_i$ and $a_j$ are the neigboring vertices, $\epsilon$
is the mode eigenvalue, and $N$ is the number of the vertices of
the complex. The normalized modes of the second type, or anomalous modes,
for certain neighboring pairs of vertices whose number is
comparable with the
total vertices number of the complex satisfy the conditions
\begin{gather}
|\psi^{{\cal A}}_i-\psi^{{\cal A}}_j|\sim|\psi^{{\cal
A}}_j|\sim N^{-1/2}.
\label{ancor360}
\end{gather}
Relations (\ref{ancor360}) mean that anomalous modes generally
change in spurts from a vertex to a neighboring vertex.
Hence, the derivatives $\partial_a\psi^{{\cal A}}_i$
of anomalous modes also generally
change in spurts from a vertex to a neighboring vertex.
Otherwise, Eqs. (\ref{ancor310}) and (\ref{ancor330})
would provide the equation
$i\gamma^a\partial_a\psi^{{\cal
A}}=\epsilon\,\psi^{{\cal A}}$, where the left-hand side is continuous,
while the right-hand side is discontinuous. Moreover, the derivatives of
anomalous modes in various directions are incommensurable.

Further, we will show that the effective equation in the continuum limit
that describes anomalous modes has the form
\begin{gather}
i\alpha^a(x)\partial_ag^{{\cal A}}(x)=\epsilon\,g^{{\cal
A}}(x),
\label{ancor370}
\end{gather}
where $\alpha^a(x)$ are random functions in the sense indicated below.
Indeed, let $\psi^{{\cal A}(0)}_{s\,j}, s=1,\,2,\,\ldots$ be
a complete set of zeroth anomalous modes\footnote{Zeroth
modes satisfy Eq. (\ref{ancor310}) with zero right-hand side.}.
Any linear combination of zeroth modes is evidently also a zeroth mode.
For this reason, we seek a soft anomalous mode "growing"
from zeroth anomalous modes in the form
$\psi^{{\cal A}}_j=\sum_sg^{{\cal
A}}_{s\,j}\psi^{{\cal A}(0)}_{s\,j}$,
where the numerical field $g^{{\cal A}}_{s\,j}$ is slowly varying.
To derive an equation for the field $g^{{\cal A}}_{s\,j}$,
we substitute the expression for a soft anomalous mode into
Eq. (\ref{ancor310}):
\begin{eqnarray}
\frac{i}{4}\sum_a\sum_{j(i)}S^a_{ij}\gamma^a\sum_{s'}\left[\left(g^{{\cal
A}}_{s'\,j}-g^{{\cal A}}_{s'\,i}\right)+g^{{\cal
A}}_{s'\,i}\right] \psi^{{\cal A}(0)}_{s'\,j}=
\nonumber \\
=\frac{i}{4}\sum_a\sum_{j(i)}S^a_{ij}\gamma^a\sum_{s'}\psi^{{\cal
A}(0)}_{s'\,j}\left(g^{{\cal A}}_{s'\,j}-g^{{\cal
A}}_{s'\,i}\right)=
\nonumber \\
=\epsilon\,v_i \sum_{s'}g^{{\cal
A}}_{s'\,i}\psi^{{\cal A}(0)}_{s'\,i}.
 \label{ancor380}
\end{eqnarray}
The first equality in Eq. (\ref{ancor380}) follows from
the fact that $\psi^{{\cal A}(0)}_{s\,j}, s=1,\,2,\,\ldots$
are zeroth modes. According to the second equality
in Eq. (\ref{ancor380}), the eigenvalue $\epsilon$ for slowly
varying fields $g^{{\cal A}}_{s\,i}$ can be arbitrarily small.
In this case, Eq. (\ref{ancor380}) is reduced to the form
\begin{eqnarray}
\frac{i}{4v_i}\sum_{s'}\left[\sum_a\sum_{j(i)}\left(\opsi^{{\cal
A}(0)}_{s\,i}S^a_{ij}e^b_{ij}\gamma^a\psi^{{\cal
A}(0)}_{s'\,j}\right)\right]\partial_bg^{{\cal A}}_{s'\,i}=
\nonumber \\
=\epsilon\left[\sum_{s'}\opsi^{{\cal A}(0)}_{s\,i}\psi^{{\cal
A}(0)}_{s'\,i}\right]g^{{\cal A}}_{s'\,i}.
\nonumber
\end{eqnarray}
The multiplication of this equation by the matrix $\left[\opsi^{{\cal
A}(0)}_{s\,i}\psi^{{\cal A}(0)}_{s'\,i}\right]^{-1}$,
reduces it to Eq. (\ref{ancor370}), where
\begin{eqnarray}
\alpha^a_{s\,s'}=\frac{1}{4v_i}\sum_{s''}\left[\opsi^{{\cal
A}(0)}_{s\,i}\psi^{{\cal
A}(0)}_{s''\,i}\right]^{-1}
\nonumber \\
\times\left[\sum_b\sum_{j(i)}
\left(\opsi^{{\cal
A}(0)}_{s''\,i}S^b_{ij}e^a_{ij}\gamma^b\psi^{{\cal
A}(0)}_{s'\,j}\right)\right].
 \label{ancor390}
\end{eqnarray}
Thus, the operator on the right-hand side of Eq. (\ref{ancor370})
is a first-order differential operator with variable matrix
coefficients that acts on a multicomponent (or vector) function
$g_s(x), \ s=1,\,2,\ldots$.

Quantities (\ref{ancor390}) are irregular functions that depend strongly
and locally on the positions of the vertices of the complex.
This fact is demonstrated by the evident formulas in the two-dimensional
theory (see \cite{4}). I give here only qualitative description of
the quantities (\ref{ancor390}) and their properties.

As it is known, the zeroth anomalous modes look like
\begin{eqnarray}
\psi^{{\cal A}(0)}_{s\,j}=\exp\left(\frac{2\pi in_s(j)}{n_s}\right),
 \label{ancor3100}
\end{eqnarray}
where $n_s$ and $n_s(j)$ are whole numbers, $n_s>1$, $n_s(j)=0,\,
\pm 1,\,\ldots\,,\,\pm(n_s-1)$, and $n_s(j_1)\neq n_s(j_2)$ for
the neighboring vertices $a_{j_1}$ and $a_{j_2}$. From here
it follows that
\begin{eqnarray}
\psi^{{\cal A}(0)}_{s\,j_1}\neq \psi^{{\cal A}(0)}_{s\,j_2}
\label{ancor3110}
\end{eqnarray}
for the neighboring vertices $a_{j_1}$ and $a_{j_2}$.
In consequence of Eqs. (\ref{ancor3100}) and (\ref{ancor3110})
the sums in the second
square brackets in (\ref{ancor390})
become very irregular \footnote{Note that for the normal
zeroth modes due to the firs of Eqs. (\ref{ancor330})
the quantity (\ref{ancor390}) is equal to $\gamma^a$.}.
Even the signs of these sums and the oriented volume $v_i$
in (\ref{ancor330}) are not correlated.

Thus, the components $\alpha^a(x)$ in Eq. (\ref{ancor370}) are
discontinuous functions of $x$ and they depend strongly and locally
on the integration variables $\{e_{\cE}\}$ in the integral (\ref{discr90}).
It is important that the variables $\{e^a_{ij}\}$ in
integral (\ref{discr90}) in the quasi-classical phase can
independently vary over a wide range without a change in the action
(see the text between Eqs. (\ref{discr160}) and (\ref{discr170})).

Further I assume that
\begin{eqnarray}
\alpha^a(x)=\beta^a(x)+\lambda\rho^a,
\label{ancor3120}
\end{eqnarray}
where  $\rho^a_{s\,s'}$ is the constant matrix, $\lambda$ is
the small numerical parameter, and all odd powers of
$\beta(x)$ in the integrand in Eq. (\ref{discr90}) are equal to zero:
\begin{eqnarray}
\big{\langle}\,|\beta(x_1)\,\ldots\,\beta(x_{2n+1})\,|\,\big{\rangle}_{e_{\cE}}=0,
\quad n=0,\,1,\,\ldots\,.
\label{ancor3130}
\end{eqnarray}
Here the matrix product of matrices $\beta$ is averaged.
The points $x_1,\,\ldots\,,\,x_{2n+1}$ can partially or completely
coincide with each other. Averaging in Eq. (\ref{ancor3130}) is performed
by means of functional integral (\ref{discr90}). The subscript
$\{e_{\cE}\}$ indicates that integration is performed only with
respect to the variables $\{e_{\cE}\}$
\footnote{Owing to the constraints (\ref{discr170}), the
variables $\{\Omega_{\cE}\}$ are insignificant here.}. We also
assume that the parameter $\lambda$ is small so that the
expansion in this parameter is meaningful. The last assumption is in
agreement with said above.

When the problem of the $S$ matrix (which is meaningful only in
the continuum limit) is solved in the theory under consideration,
all $S$ matrix elements mast be averaged over the variables
$\{e_{\cE}\}$ before the calculation of probabilities and cross sections.
Therefore, if the vertices describing the interaction are universal
(i.e., independent of the microstructure of the lattice), the propagators of the
matter fields in such a theory must be averaged over the variables
$\{e_{\cE}\}$. Indeed, in this case, the diagrammatic technique can be
obtained in this case as a result of the expansion in a functional differential
operator acting on the transition amolitude for the matter fields in the
quadratic ("free") approximation against the background of the external
field sources. In this case, the square transition amplitude must be
averaged with a weight over the variables $\{e_{\cE}\}$
before the expansion in interaction.

So, it is necessary to obtain the propagator
\begin{eqnarray}
\langle\psi^{{\cal A}}\opsi^{{\cal A}}\rangle_{e_{\cE}}=
\psi^{{\cal
A}(0)}\big{\langle}\big(-i\alpha^a\partial_a\big)^{-1}\big{\rangle}_{e_{\cE}}\opsi^{{\cal
A}(0)},
\label{ancor3140}
\end{eqnarray}
that describes the propagation of anomalous modes. Thus, the problem
is reduced to the study of the Green's function that corresponds to the
operator on the left-hand side of Eq. (\ref{ancor370}) and is
averaged over the variables $\{e_{\cE}\}$ by taking into account conditions
(\ref{ancor3130}): \footnote{It is necessary to point to a fundamental
difference of this computational procedure from that used in problem of the
localization of particles in random potentials. The physical difference between
these situations is that the parameters characterizing the randomness
of a potential in the latter case are not dynamical degrees of freedom,
whereas the positions of the vertices of the complex are dynamical variables.}:
\begin{eqnarray}
\big{\langle}\big(-i\alpha^a\partial_a\big)^{-1}\big{\rangle}_{e_{\cE}}.
\label{ancor3150}
\end{eqnarray}

{\bf {4.}} We first consider the case $\lambda=0$.
To perform the necessary averaging in Eq. (\ref{ancor3140}),
we use the well-known formula
\begin{eqnarray}
{\cal P}\frac{1}{\tau}=\frac{1}{2i}\int_0^{+\infty}\d
s\left[e^{is\tau}-e^{-is\tau}\right].
\label{aver410}
\end{eqnarray}
This formula implies that $\tau\neq 0$ or
$\tau^{-1}\neq\pm\infty$. Using Eq. (\ref{aver410}), we represent
operator (\ref{ancor3150}) (for the case $\lambda=0$ before
averaging over the variables $\{e_{\cE}\}$) in the form
\begin{eqnarray}
\big{\langle}x\big|\big(-i\beta^a\partial_a\big)^{-1}\big|y\big{\rangle}=
\nonumber \\
=\frac{1}{2i}\int_0^{+\infty}\d
s\,\big{\langle}x\big|\left[e^{s\beta^a\partial_a}-
e^{-s\beta^a\partial_a}\right]\big|y\big{\rangle}.
\label{aver420}
\end{eqnarray}
This relation is meaningful for $x\neq y$, because its left-hand side
is finite in this case. However, representation (\ref{aver420})
is meaningless for $x=y$. Then, let us average the terms
in the square brackets in Eq. (\ref{aver420}) over the variables
$\{e_{\cE}\}$. In this case, it is only important that, owing
to Eq. (\ref{ancor3130}), the result of such averaging of each
of these terms is a function of $(\pm s)^2=s^2$ but not of
$(\pm s)$. Therefore, the terms in the square brackets in
Eq. (\ref{aver420}) cancel each other after the averaging
over the variables $\{e_{\cE}\}$ and thereby the propagator
describing the propagation of anomalous modes is proportional to
the $\delta$ function of the variable $x$:
\begin{eqnarray}
\left\langle\big{\langle}x\big|\big(-i\beta^a\partial_a\big)^{-1}
\big|y\big{\rangle}\right\rangle_{e_{\cE}}\sim
a\,\delta^{(4)}(x-y).
\label{aver430}
\end{eqnarray}
I remind the reader that $a$ is a certain constant
that has the length dimension and is of the order of
the effective lattice step.

Now let's estimate quantity (\ref{ancor3150}) for nonzero
$\lambda$ by taking into account that it can be expanded
in the parameter $\lambda$. This problem is easily solved by using
field theory methods. Figure 1 shows the sum of the diagrams
that corresponds to quantity (\ref{ancor3150}). The solid line and
cross correspond to unperturbed propagator (\ref{aver430}) and
the operator $i\lambda\rho^a\partial_a$, respectively.
Then using the conventional diagrammatic technique rules, one can
represent quantity (\ref{ancor3150}) in the form of the
series of diagrams shown in Fig. 1.

\addtocounter{figure}{0}
{\psfrag{Ph1}{\kern-5pt\lower-1pt\hbox{\large $\phi_1$}}
\psfrag{Ph2}{\kern0pt\lower0pt\hbox{\large $\phi_2$}}
\psfrag{Ph3}{\kern0pt\lower0pt\hbox{\large $\phi_3$}}
\psfrag{Ia}{\kern0pt\lower0.5pt\hbox{\large $a$}}
\psfrag{Ib}{\kern0pt\lower0.5pt\hbox{\large $b$}}
\psfrag{Ic}{\kern0pt\lower0.5pt\hbox{\large $c$}}
\begin{figure}[tbp]
 \includegraphics[width=0.20\textwidth]{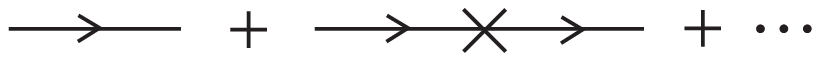}
\caption{} \label{One}
 \end{figure}}

As a result, the $\delta$ function of the free propagator
is "smeared" and the desired propagator describing the propagation
of anomalous modes appears to be exponentially decreasing
in the $x$ space according to the Eq. (\ref{introduction10}).

At the same time, according to Eq. (\ref{ancor340}), the free
propagators describing the propagation of normal modes
behave according to the power law in $x$ and momentum space.

{\bf {5.}} We briefly summarize the conclusions as follows.

(i) On complexes that formally allow state doubling (of any fields), this
phenomenon is really absent because normal and anomalous
modes propagate differently and separately in spacetime.
Normal modes (in the long-wavelength limit) have definite
energy and momentum. On the contrary, anomalous modes cannot
have definite energy and momentum and their propagators decrease
exponentially in spacetime at scales comparable with the
characteristic lattice scale.

(ii) Let's give some general considerations
concerning fermion (Wilson) doubling due to the above result.

Let us assume that one Weyl field is introduced on the
lattice by introducing the projection operator
$(1/2)(1\pm\gamma^5)$ into the fermion action in
Eq. (\ref{discr40}). Since the lattice fermion action
is invariant under global $\gamma^5$ transformations and
the lattice fermion measure is invariant under local
$\gamma^5$ transformations, the total fermion current is
strictly conserved. However, this does not mean that each
of the currents of normal and anomalous modes is conserved
separately. On the contrary, it has been well known for a
long time that, when $S$ matrix elements are calculated by
using causal Feynman propagators, the Weyl-field current
is not conserved due to the gauge anomaly. This phenomenon is
interpreted in the theory under consideration as the mutual
transfer of the axial charge between normal and anomalous modes.
Here it is shown that this phenomenon can occur only on the
lattice scales.

I also note that the theory under consideration does not
exclude the case where the axial chargees of normal and
anomalous modes are conserved separately (see \cite{3}).
This means that the axial anomaly is absent. Such a regime
can be realized only in problems where the use of Feynman
propagators is incorrect, for example, at the universe
inflation stage when the problem of the $S$ matrix is meaningless.

\begin{acknowledgments}

This work was
supported by RFBR No. 04-02-16970-a.

\end{acknowledgments}


\begin{thebibliography}{99}


\bibitem{1}
S.N. Vergeles, JETP, {\bf{93}}, 926 (2001).

\bibitem{2}
S.N. Vergeles, JETP Letters, {\bf{76}}, 1, (2002).

\bibitem{3}
S.N. Vergeles, JETP, {\bf{97}}, 1075 (2003).

\bibitem{4}
S.N. Vergeles, JETP Letters, {\bf{82}}, 617 (2005).

\bibitem{5}
S.N. Vergeles, Nucl. Phys. {\bf{B 735}}, 172 (2006); hep-th/0512137.

\bibitem{6}
P. Menotti and A. Pelissetto, Phys. Rev. {\bf{D 35}}, 1194 (1987).

\bibitem{7}
K. G. Wilson, Erice lectures notes (1975);
J. Kogut, and L. Susskind, Phys. Rev. {\bf {D 11}}, 393 (1975);
L. Susskind, Phys. Rev. {\bf {D 16}}, 3031 (1977);
Martin Luscher, E-print archives hep-th/0102028;
H. B. Nielsen and M. Ninomiya, Nucl. Phys. {\bf
{B 185}}, 20 (1981); Nucl. Phys. {\bf {B 193}}, 173 (1981).



\end{thebibliography}
\end{document}